\documentclass[aps,prb,twocolumn,amsmath,amssymb,superscriptaddress]{revtex4-1}

\usepackage{xspace}
\usepackage{graphicx}
\usepackage{dcolumn}
\usepackage{bm}
\usepackage{color}

\newcommand{\cmi}{$\mathrm{cm}^{-1}$\xspace}
\newcommand{\BIO}{$\mathrm{Ba}_{2}\mathrm{In}_{2}\mathrm{O}_{5}$\xspace}
\newcommand{\BIOH}{$\mathrm{Ba}\mathrm{In}\mathrm{O}_{3}\mathrm{H}$\xspace}

\begin{document}

\title{Structure and dehydration mechanism of the proton conducting oxide \boldmath{Ba$_{2}$In$_{2}$O$_{5}$(H$_{2}$O)$_{x}$}}

\author{Johan Bielecki}
\email{johanb@xray.bmc.uu.se}
\affiliation{Department of Applied Physics, Chalmers University of Technology, SE-41296 G{\"o}teborg, Sweden}
\affiliation{Department of Cell and Molecular Biology, Uppsala University, Box 596, SE-75124 Uppsala, Sweden}
\author{Stewart F. Parker}
\affiliation{ISIS Facility, STFC Rutherford Appleton Laboratory, Chilton, Didcot, Oxon OX11 0QX UK}
\author{Lars B\"{o}rjesson}
\affiliation{Department of Applied Physics, Chalmers University of Technology, SE-41296 G{\"o}teborg, Sweden}
\author{Maths Karlsson}
\email{maths.karlsson@chalmers.se}
\affiliation{Department of Applied Physics, Chalmers University of Technology, SE-41296 G{\"o}teborg, Sweden}
\date{\today}

\begin{abstract}

The structure and dehydration mechanism of the proton conducting oxide Ba$_{2}$In$_{2}$O$_{5}$(H$_{2}$O)$_{x}$ are investigated by means of variable temperature Raman spectroscopy together with inelastic neutron scattering.
At room temperature, Ba$_{2}$In$_{2}$O$_{5}$(H$_{2}$O)$_{x}$ is found to be fully hydrated ($x=1$) and to have a perovskite-like structure, which dehydrates gradually with increasing temperature and at around 600~$^{\circ}$C the material is essentially completely dehydrated ($x=0$).
The dehydrated material exhibits a brownmillerite structure, which is featured by alternating layers of InO$_{6}$ octahedra and InO$_{4}$ tetrahedra.
The transition from a perovskite-like to a brownmillerite-like structure is featured by a hydrated-to-intermediate phase transition at \emph{ca.} 370~$^{\circ}$C.
The structure of the intermediate phase is similar to the structure of the fully dehydrated material, but with the difference that it exhibits a non-centrosymmetric distortion of the InO$_{6}$ octahedra not present in the latter. 
For temperatures below the hydrated-to-intermediate phase transition, dehydration is featured by the release of protons confined to the layers of InO$_{4}$ tetrahedra, whereas above the transition also protons bound to oxygens of the layers of InO$_{6}$ are released. 
Finally, we found that the O-H stretch region of the vibrational spectra is not consistent with a single-phase spectrum, but is in agreement with the superposition of spectra associated with two different proton configurations.
The relative contributions of the two proton configurations depend on how the sample is hydrated.

\end{abstract}


\maketitle

\section{Introduction}

\begin{figure*}[tbh]
\begin{center}
\includegraphics[width=0.93\linewidth]{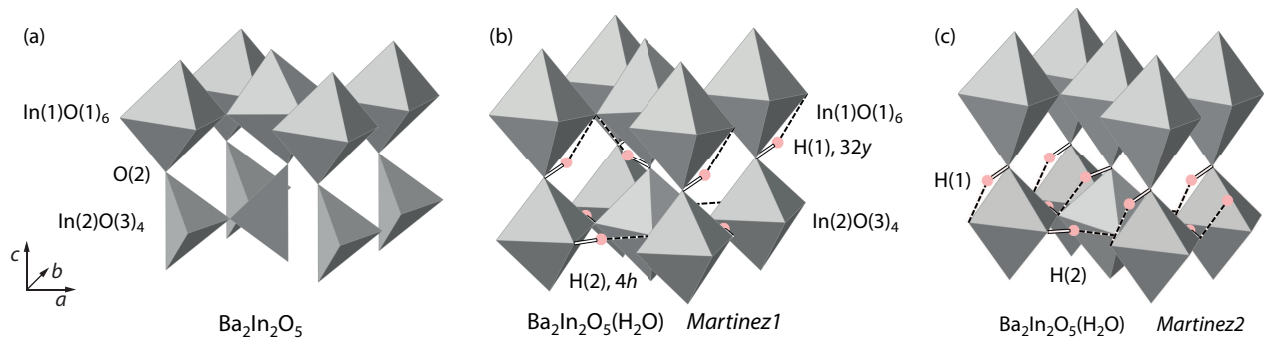}
\caption{
Schematic illustration of the structure of \BIO (a) and the two lowest-energy proton configurations in \BIOH (b-c), according to Martinez \emph{et al.}\cite{Martinez}. 
InO polyhedra with oxygens located at the vertices are depicted in grey, and H by pink spheres. 
Covalent O-H bonds are indicated by solid grey lines, and hydrogen bonds are indicated by dashed lines.
Ba atoms are omitted for simplicity. 
We note that even though \BIOH is solely built up of InO$_{6}$ octahedra, we will keep the ``tetrahedral'' and ``octahedral'' nomenclature to distinguish the two layers from each other. \vspace{-4mm}
}
\label{fig:Structure}
\end{center}
\end{figure*}

Proton conducting oxides, which may be classified according to their structure type, are currently accumulating considerable attention due to their significant potential as efficient electrolytes in next-generation, intermediate-temperature ($\approx$200--500~$^{\circ}$C) solid oxide fuel-cell (SOFC) technology based on proton conducting electrolytes.\cite{KRE03,MAL10}
Amongst the most promising materials, is the brownmillerite structured oxide \BIO, which may be described as an oxygen deficient variant of the perovskite structure, with alternating layers of InO$_{6}$ octahedra and InO$_{4}$ tetrahedra, with no orientational order between successive layers.\cite{BIE14}
 As shown in Fig.~\ref{fig:Structure}(a), the octahedral layers contain the In(1) and O(1) atomic positions and the tetrahedral layers contain the In(2) and O(3) atomic positions, with the two types of layers bridged by the apical oxygens, denoted O(2).

Like most other oxygen deficient oxides, \BIO transforms upon hydration into a hydrogen containing, proton conducting, material.
Hydration is generally carried out by heat treatment in a humid atmosphere, a process during which the water molecules in the gaseous phase dissociate into hydroxyl groups (-OH$^{-}$) and protons (H$^{+}$) on the surface of the sample. 
The -OH$^{-}$ groups then stick to nearby oxygen vacancies, whilst the remaining protons bind to lattice oxygens of the oxide host lattice. 
The protons, however, are not stuck to any particular oxygen atoms, but are  free to move from one oxygen to another and, with time, they will therefore diffuse into the bulk of the material. 
At the same time as protons diffuse into the bulk, the counter diffusion of oxygen vacancies from the bulk to the surface allows the dissociation of other water molecules on the surface of the sample.
This leads to an increase of the proton concentration in the material and so it is believed that the process continues until the (bulk) oxygen vacancies are filled, leading, ideally, to a material of the form \BIOH. 

The structure of \BIOH is not a brownmillerite, but may be described as a perovskite-like structure with successive, distinctly different, layers of InO$_{6}$ octahedra running along the $c$-direction of an orthorhombic unit cell, \emph{cf.} Fig.~\ref{fig:Structure} (b)--(c). 
The orthorhombic arrangement can be expected to be due to proton ordering, as opposed to protons being randomly distributed over the oxide host lattice.
Neutron diffraction analysis showed that the average structure contains two different proton sites, one which refers to the midpoint between O(1) atoms within the octahedral layer and the other one which refers to a position in the plane formed by the apical O(2) oxygens, described by the 2$c$ and 16$l$ Wyckoff positions, respectively.\cite{Jayaraman}
Using these results as a starting point for structural optimizations by means of first-principles calculations, Martinez \emph{et al.}\cite{Martinez} and Dervi\c{s}o\u{g}lu \emph{et al.} \cite{Riza2015} both investigated the possible local proton configurations and found that the 16$l$ protons are, in a more realistic proton arrangement, described by the 32$y$ position and that the 2$c$ protons are described by the 4$h$ position, where 4$h$ and 32$y$ represent local deviations from the average 2$c$ and 16$l$ positions.\cite{Martinez}
Specifically, the 4$h$ position refers to protons which we here denote as H(2) and which are bonded to oxygens of the tetrahedral layers, O(3), whereas the 32$y$ position refers to protons which we denote as H(1) and which are bonded to the apical oxygens, O(2), \emph{cf.} Fig.~\ref{fig:Structure}(b).
Both of the theoretical studies found two local structures (proton configurations), labeled \emph{Martinez1} and \emph{Martinez2}, as shown in Fig. \ref{fig:Structure}(b) and (c), with lower energies compared to a range of other proton configurations also considered in the structural optimizations.
The two studies do not, however, agree on the ground-state structure: whereas Martinez \emph{et al.} \cite{Martinez} assigned the ground-state structure to the \emph{Martinez1} proton configuration, Dervi\c{s}o\u{g}lu \emph{et al.} \cite{Riza2015} found that \emph{Martinez2} was of lowest energy.  
The two local structures are conceptually similar, with equally many protons in the $4h$ and $32y$ positions, respectively, and where the only difference between them relates to the hydrogen-bond pattern of the $32y$ protons. 
In the \emph{Martinez1} structure, the $32y$ protons are hydrogen bonded towards the O(1) layer, whereas in the \emph{Martinez2} structure the $32y$ protons are hydrogen bonded to the O(3) oxygens. 
Recently, it was shown that the hydrogen bonding of $32y$ protons in the \emph{Martinez1} structure has the effect of pulling the O(1) oxygen towards the H(1) site and thereby gives rise to a long-range non-centrosymmetric distortion of the In(1)O$_6$ octahedra.\cite{BIE14}
Further, Dervi\c{s}o\u{g}lu \emph{et al.} \cite{Riza2015} measured the $^1$H NMR spectra of \BIOH, which suggested the presence of three distinct proton positions in the structure. 
The three positions correspond to one position within the O(3) layer, and two positions within the O(2) layer that hydrogen bonds to either the O(3) or O(1) layer, respectively.
First-principles calculations could reproduce the $^1$H NMR experiments by including four low energy proton configurations within the material.\cite{Riza2015} 
Each of these configurations, labeled I, J K, and L, is a specific combination of proton occupations on the three positions mentioned above.\cite{Riza2015} 
Similarities in hydrogen bond patterns and crystal distortions make it possible to associate, with regards to vibrational fingerprints, I and K as ''\emph{Martinez1}-like'' proton configurations, whereas J and L can be regarded as ''\emph{Martinez2}-like''.

While the structures of the fully dehydrated and fully hydrated structures have emerged recently,\cite{BIE14} little is known about the structure for intermediate proton loadings, \emph{i.e.} for partially hydrated structures, and in particular how that may influence the material's proton conducting properties. 
In this context, it has been suggested recently that the full occupation of H(2) protons on the 4$h$ site may hinder the diffusion of protons within the In(2)-O(3) plane containing the nearest oxygen neighbors to which the H(2) protons form strong hydrogen bonds, and therefore that the proton conductivity may be governed instead by the more weakly hydrogen bonded H(1) protons on the 32$y$ site.\cite{BIE14} 
However, upon dehydration with increasing temperature it might be that the diffusivity of H(2) protons increases at a rate that is a function of the H(2) occupancy, and if so, the question is whether there is an optimum occupancy? 
Such information is not only of purely academic interest, but can be expected to help significantly in the development of new, more highly proton conducting oxide systems; this is in turn critical for the development and eventually commercialization of intermediate-temperature SOFC technology based on proton conducting electrolytes.
For this reason, we  investigate here the dehydration mechanism and, in particular, the structure for intermediate proton loadings of Ba$_{2}$In$_{2}$O$_{5}$(H$_{2}$O)$_{x}$.
The investigations are performed by means of variable temperature Raman spectroscopy together with inelastic neutron scattering (INS), which provide detailed information about the local structures present as well as their vibrational fingerprints. 
We also discuss our structural results in terms of possible proton conduction mechanisms. 

\section{Experimental}

\subsection{Sample preparation}
A powder sample of Ba$_2$In$_2$O$_5$ was prepared by solid state sintering by mixing stoichiometric amounts of the starting reactants (BaCO$_{3}$ and In$_{2}$O$_{3}$), with the sintering process divided into three treatments: 1000~$^{\circ}$C for 8 h, 1200~$^{\circ}$C for 72 h, and at 1325~$^{\circ}$C for 48 h, with intermediate cooling, grinding and compacting of pellets between each heat treatment. 
The as-sintered Ba$_{2}$In$_{2}$O$_{5}$ powder was annealed in vacuum at high temperature ($\approx$600~$^{\circ}$C) in order to remove any protons that the sample may have taken up during its exposure to ambient conditions; this sample is referred to as dehydrated and exhibited essentially the same spectrum as a hydrated sample, \BIOH, after heating to 600~$^{\circ}$C in air.
A hydrated sample, \BIOH, was prepared by annealing a portion of the dehydrated sample at $\approx$300~$^{\circ}$C under a flow of N$_{2}$ saturated with water vapor for a period of a few days.
Thermal gravimetric analysis (TGA) indicated that this sample was found to be fully hydrated.  X-ray diffraction (XRD) patterns of both the dehydrated and hydrated samples were in agreement with the structures as reported earlier.
Further details about the sample preparation, TGA and XRD, can be found in ref. \cite{BIE14}.

\subsection{Raman spectroscopy}

The Raman spectroscopy experiments were performed in backscattering geometry using a DILOR XY800 spectrometer, equipped with a tunable $\mathrm{Ar}^+$ laser, a long working distance 40x objective,  and a liquid nitrogen cooled CCD detector.
The laser was tuned to the green 514~nm line and the laser power at the sample position was kept at 4 mW for all measurements.
A comparison of the Stokes and anti-Stokes spectra showed negligible laser heating on the sample.
All spectra were collected with linearly polarized light impinging on the sample and unpolarized light collected at the CCD, and we used three different experimental setups for our measurements.
The 35--720 \cmi range, covering the vibrational modes of the oxide host lattice, was measured in a high resolution double subtractive mode with a 800~mm focal distance.
The higher-frequency region, 2500--4000 \cmi, covering the O-H stretch vibrational modes, was measured with a single grating of 300~mm focal distance.
Variable temperature measurements were performed by measuring \emph{in-situ} at sequentially higher temperatures.
The temperature was controlled by a Linkam heating stage over the range from room temperature (RT, 20~$^\circ$C) to 600~$^\circ$C, with a small opening to prevent overpressure as the sample was dehydrated with increasing temperature.
To ensure that the spectra were measured in thermodynamic equilibrium, the sample was held for 1 h at each temperature before measuring the Raman spectra and, in addition, successive measurements at the same temperature were performed in order to rule out further dehydration after this time.
All Raman spectra have been corrected for the Bose-Einstein occupation factor and normalized to a common baseline level.

\subsection{Inelastic neutron scattering}

The INS experiment was performed on the fully hydrated sample, \BIOH, on MAPS\cite{PAR11} at 10 K with an incident energy, $E_i$, of 650 meV, with the Fermi chopper at 600 and 500 Hz.
The sample, approximately 15 grams, was loaded into an aluminium sachet and the sachet into an indium wire sealed thin-walled aluminium can, and the measuring time was about one day.

\section{Results and discussion}

\subsection{Structural variability}

\begin{figure}[t]
\begin{center}
\includegraphics[width=0.85\linewidth]{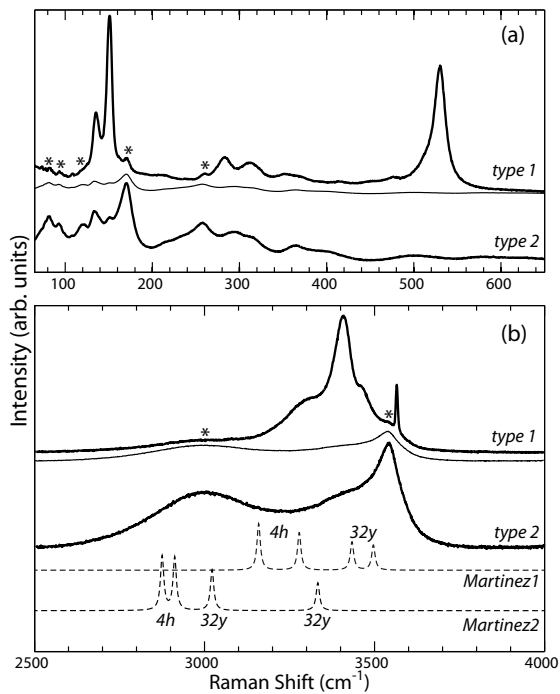}
\caption{Comparison of the Raman spectra of \BIOH from different sample batches, illustrating the spectral variations in both (a) the lattice and (b) the O-H stretch region of the vibrational spectra.
The thin solid line is a rescaled \emph{type2} spectrum that corresponds to the amount of \emph{type2} phase found in the spectrum of predominantly \emph{type1}. 
The corresponding impurity peaks are indicated by asterisks, whereas the dashed lines in (b) show the O-H stretch frequencies as determined for the two lowest-energy proton configurations as predicted by DFT calculations \cite{BIE14}.}
\label{fig:Proton}
\end{center}
\end{figure}

While investigating the RT Raman spectra of \BIOH samples from different sample batches, differing essentially in hydration conditions (hydration time and temperature), we observed significant spectral variations.
In particular, we found that these variations can be ascribed to different ratios of two distinctly different proton configurations, or phases, which are here denoted as \emph{type1} and \emph{type2}.
A comparison of the spectra of samples for which either phase is predominant (Fig.~\ref{fig:Proton}), suggests that \emph{type2} is generally characterized by a smaller number of bands related to vibrations of the oxide host lattice [Fig.~\ref{fig:Proton}(a)], and a wider O-H stretch region [Fig.~\ref{fig:Proton}(b)]. 
The former characteristic suggests a more symmetric, although not necessarily ordered, structure, whereas the latter suggests a larger variability in O-H distances in the material. 
Conversely, the narrower O-H stretch region in \emph{type1} indicates less structural variability between unit cells, whereas the presence of sharp, intense, Raman bands at around 150 and 530 \cmi, respectively, is a clear characteristic of a reduction of the symmetry of the local structure, not present in phase \emph{type2}. 

Included in Fig.~\ref{fig:Proton}(b) are also the calculated Raman spectra according to Bielecki \emph{et al.}\cite{BIE14}, for the two lowest-energy proton configurations found by Martinez \emph{et al.}\cite{Martinez} and Dervi\c{s}o\u{g}lu \emph{et al.} \cite{Riza2015}, \emph{i.e.} the proton configurations that here are called \emph{Martinez1} and \emph{Martinez2}, \emph{cf.} Fig.~\ref{fig:Structure}.
As can be seen, the \emph{Martinez1} configuration corresponds to O-H stretch modes in the relatively narrow range from 3100 to 3500 \cmi, which is in agreement with the experimental spectrum of predominantly \emph{type1}. 
In comparison, the \emph{Martinez2} proton configuration is characterized by O-H stretch modes at lower frequencies and is better in agreement with the experimental spectrum of predominantly \emph{type2}. 
The association of \emph{type1} with \emph{Martinez1}-like and \emph{type2} with \emph{Martinez2}-like structures is consistent with the low-frequency Raman spectra [Fig.~\ref{fig:Proton}(a)], where the two 150 \cmi and 530 \cmi bands, as present only in the  \emph{type1} spectrum, can be explained by the non-centrosymmetric In(1)-O$_{6}$ distortion induced by the $32y$ hydrogen bond pattern in the \emph{Martinez1} proton configuration, as mentioned above.\cite{Martinez,Riza2015}
In general, such a structural distortion activates previously inactive Raman modes; in this case an In(1) mode at 150 \cmi and an In(1)-O stretch mode at 530 \cmi. 
One should note, however, that there is a small degree of intermixing of the two phases. 
This is reflected by the thin line in both Fig.~\ref{fig:Proton}(a) and Fig.~\ref{fig:Proton}(b), which illustrate the amount of the \emph{type2} phase found in the predominating \emph{type1} sample. 
By comparing the relative contribution of the two phases to the total integrated intensity of the O-H stretch region, we estimate that the sample of predominantly \emph{type1} contains approximately 20\% of \emph{type2}. 
Lastly, note that the calculations were done in optimized, static, unit-cell geometries and hence cannot capture the unit-cell variations giving rise to the Gaussian-shaped broadenings, nor the finite vibrational lifetime giving rise to Lorentzian-shaped broadenings, in the experimental spectra. 
Thus, the calculated frequencies should be seen as indications of the frequency range expected from the different atomic positions in the experimental spectra.

\subsection{Host-lattice region of the vibrational spectra}

\label{sec:RT}

\begin{figure}[t]
\begin{center}
\includegraphics[width=0.85\columnwidth]{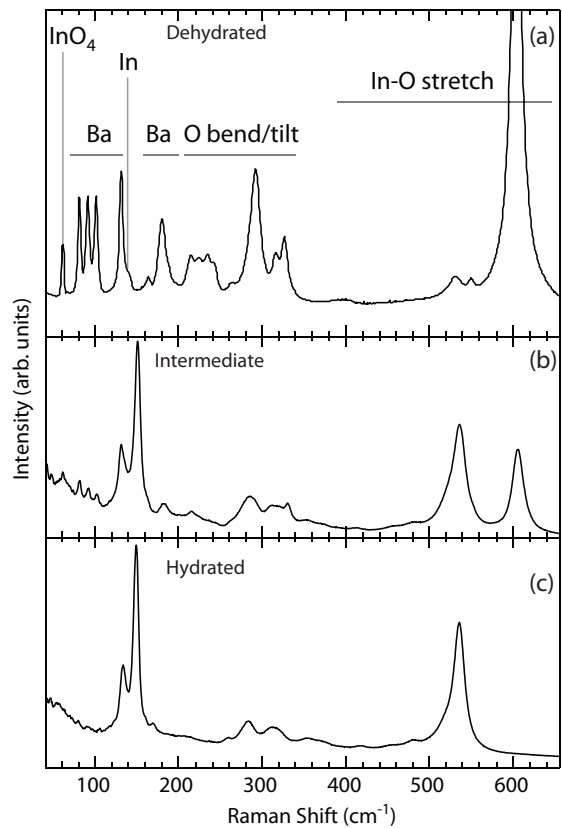}
\caption{Raman spectra of (a) dehydrated, (b) intermediate, and (c) hydrated phases of \BIO.
The atomic motions assigned to the vibrational bands in the dehydrated phase are indicated.}
\label{fig:BIO}
\end{center}
\end{figure}

In Fig.~\ref{fig:BIO} are shown the 50--650 \cmi range of the RT Raman spectra of the fully dehydrated and fully hydrated samples (\BIO and \BIOH), as well as the spectrum of an intermediate proton loading as will be discussed in detail below.
Considering first the spectrum of the dehydrated, brownmillerite structured, material [Fig.~\ref{fig:BIO}(a)], we observe several well-defined bands, in agreement with the literature.
These bands are assigned according to the following: (i)  bands below 200~\cmi relate to vibrational modes involving the heavy Ba ions, (ii) bands between 200 and 350~\cmi relate to different tilt and bend modes of the InO$_{4}$ and InO$_{6}$ moieties, and (iii) bands between 350 and 650 \cmi relate to symmetric In-O stretch modes of the same moieties.\cite{BIE14} 
The only discrepancy from this classification regards two In related bands at approximately 60 \cmi and 130 \cmi (indicated by vertical lines), respectively. 

Considering next the spectrum of the hydrated material [Fig.~\ref{fig:BIO}(c)], we observe that the spectrum changes considerably upon hydration, which is as expected since the overall structure changes from a brownmillerite to a perovskite-like structure. 
In particular, we observe that all Ba related bands, except the one at 130 \cmi, as well as the strong 600 \cmi band, which is assigned to In-O stretches of InO$_{4}$ tetrahedra, are now completely absent.
Instead, a strong band at around 530 \cmi, which is assigned to In-O stretches of InO$_{6}$ octahedra, and a band at 150 \cmi, are now observable in the spectrum.
The 150 \cmi band has previously been assigned to an In(1) related mode activated by the long-range non-centrosymmetric distortion of the In(1)O$_{6}$ octahedra, as caused by the hydrogen bonding between H(1) protons and O(1) oxygens, which is a fingerprint of the \emph{Martinez1} proton configuration.\cite{BIE14}
Further information about the non-centrosymmetric In(1)O$_6$ distortion can be found in the linewidths of the 150 and 530 \cmi bands as a function of temperature, as we shall see later but first we discuss the overall spectral changes with temperature.

\begin{figure*}[t]
\begin{center}
\includegraphics[width=1.8\columnwidth]{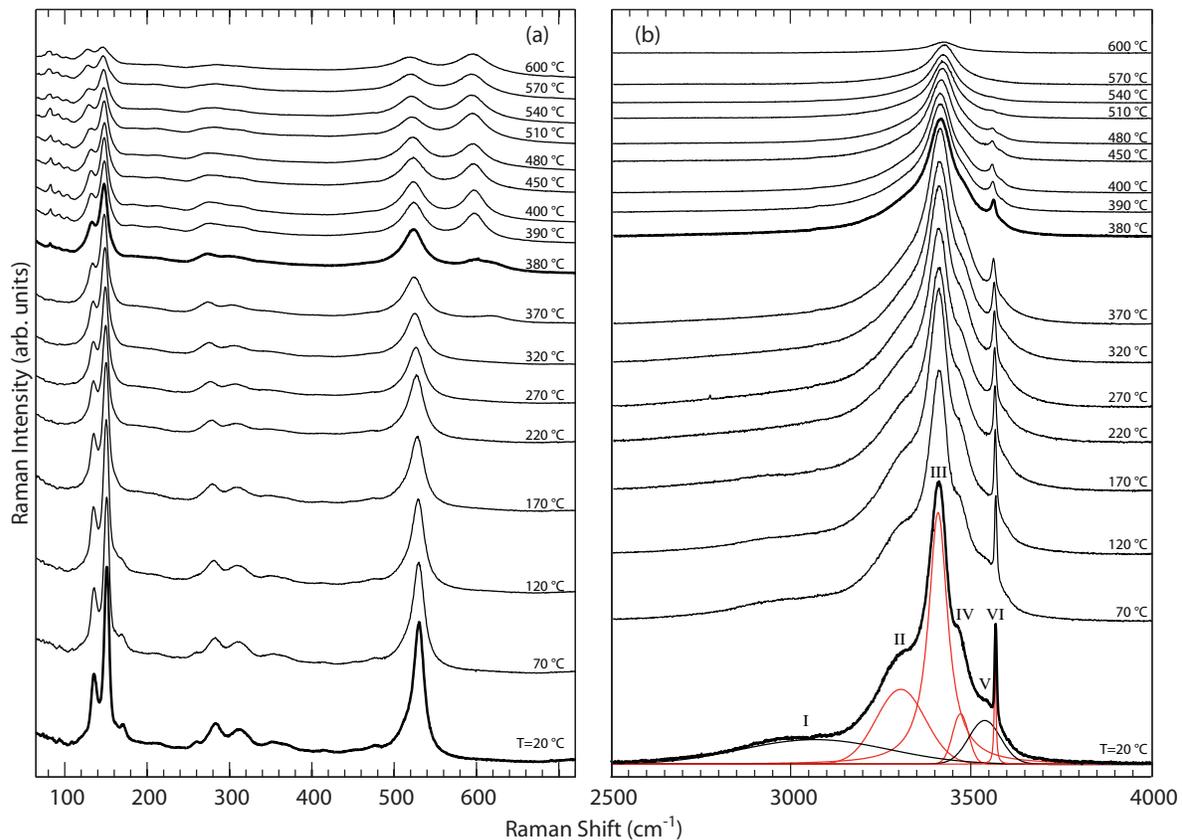}
\caption{(Color online) Variable temperature Raman spectra measured in (a) the lattice and (b) the O-H stretch frequency region, respectively.
The spectra distinguished with thick lines indicate the fully hydrated sample and the appearance of the intermediate phase.
Included in (b) is the peak-fitted components of the O-H stretch spectrum, where the bands II, III, IV and VI (in red color) are suggested to relate to the \emph{Martinez1} structure and the bands I and V (in black color) are suggested to relate to the \emph{Martinez2} structure. \vspace{-3mm}
}
\label{fig:Tstudy}
\end{center}
\end{figure*}

Figure~\ref{fig:Tstudy} shows the Raman spectra measured upon increasing the temperature from 20 to 600 $^\circ$C. 
For the 50--720 \cmi range of the spectra [Fig.~\ref{fig:Tstudy}(a)], which relates to the vibrational dynamics of the oxide lattice, we observe a general broadening of all bands as a function of increasing temperature from 20 to 370~$^\circ$C.
At a temperature of  370--380~$^\circ$C the spectrum changes more markedly. 
Most noticeable is the appearance of new, rather strong, bands, at approximately 60, 82, 92, 102, 180 and 620 \cmi, as well as of weaker bands in the range 215--243 \cmi, suggesting a structural phase transition away from the structure of the (fully) hydrated material.
In this context, the 60 \cmi and 620 \cmi bands are identified as tilt motions and symmetric In-O stretches of InO$_4$ tetrahedra, respectively.\cite{BIE14}
The appearance of these bands is in agreement with the concomitant transformation of InO$_6$ octahedra to InO$_4$ tetrahedra as the sample is dehydrated with increasing temperature.\cite{BIE14}
The other bands are related to vibrations involving mainly the Ba ions (82, 92, 102, and 180 \cmi) and oxygen ions (215--243 \cmi), respectively, further reflecting the structural change.
Upon further temperature increase (from 370~$^\circ$C to 600~$^\circ$C), the spectrum changes smoothly towards the shape of the spectrum for the dehydrated material.
In particular, the symmetric In-O stretch band at 620 \cmi band downshifts gradually with increasing temperature to reach a position of 595 \cmi at 600~$^\circ$C.

\begin{figure*}
\begin{center}
\includegraphics[width=0.8\linewidth]{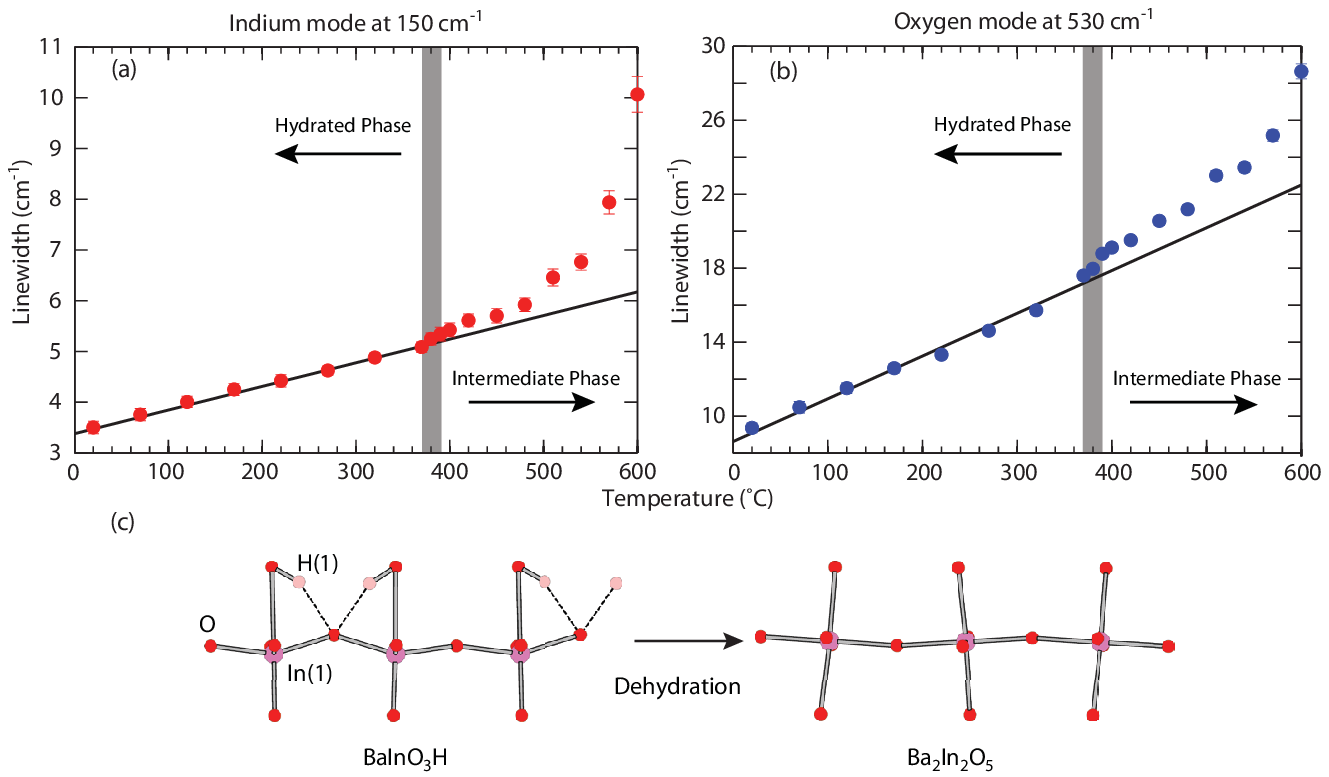}
\caption{
Temperature dependence of the spectral linewidth (FWHM) of (a) the indium mode at 150 \cmi and (b) the oxygen mode at 530 \cmi. 
The anomalous increase in linewidth above the hydrated-to-intermediate phase transition is attributed to a gradual decoherence of the non-centrosymmetric distortion of the In(1)O$_{6}$ octahedra as H(1) protons are released; this is depicted schematically in (c).\vspace{-3mm}
}
\label{fig:LineWidth}
\end{center}
\end{figure*}

Let us now turn to the temperature dependence of the linewidths of the two In(1) related bands at 150 and 530 \cmi, which is associated with the non-centrosymmetric In(1)O$_6$ distortion in \BIOH.
 The thermal linewidth broadening is given by the Klemens model, which takes into account the anharmonic decay of one optical phonon into two acoustic phonons.\cite{Klemens} 
 By this process the linewidth, $\Gamma$, increases with temperature $T$ according to $\Gamma(T) \approx \Gamma(0)\left[1+2/(\exp{(\hbar \omega_0/2k_BT)}-1)\right]$, where  $\omega_0$ is the frequency of the optical phonon. Deviations from this rule are a sign of additional processes that decrease the phonon lifetime $\tau$ ($\tau \approx 1/\Gamma$) and broadens the vibrational linewidth. 
 Such broadening commonly arises from increased disorder, and consequently anharmonicity, of the atomic species involved in the vibration at hand.\cite{Perry,Loudon}
 
In Fig.~\ref{fig:LineWidth}(a) is shown the temperature evolution of the 150 and 530 \cmi linewidths, together with fits to the Klemens model (solid lines).
As can be seen, the measured linewidths agree well with the Klemens model until a temperature of \emph{ca.} 370 $^\circ$C is reached, indicating no loss of coherence in the InO$_6$ distortion below 370 $^\circ$C. 
Above 370 $^\circ$C, however, both modes show an anomalous increase of $\Gamma$ with increasing temperature. 
This is a clear indication of decoherence in the In(1)O$_6$ distortion, which we interpret as due to the gradual release of H(1) hydrogens above 370 $^\circ$C, see Fig. \ref{fig:LineWidth}(c). 
Thus, the intermediate structure is distinctly different from the dehydrated structure in that, even though the crystal structure approaches the dehydrated structure upon dehydration, the non-centrosymmetric InO$_6$ distortion is still present. 
This picture is also supported by the temperature trends in the O-H stretch peak positions (as will be discussed later), where only the bands coupled to the H(2)-O(3) layer are affected by the transition.

\subsection{O-H stretch region of the vibrational spectra}

The gradual dehydration upon increasing temperature is consistent with the spectral changes of the O-H stretch region [Fig.~\ref{fig:Tstudy}(b)], which reflects a change in the local coordination of protons in the material.
In particular, one should note that the frequency of an O-H stretch mode is very sensitive to the degree of hydrogen bonding the proton may experience towards a neighboring oxygen and that such a hydrogen-bonding interaction generally softens the mode. 
Analysis of the O-H stretch band/s provides therefore a spectroscopic means not only to identify, but also to distinguish between different proton sites in the structure.
The very broad, asymmetric, O-H stretch band for \BIOH hence suggests that several different proton sites are present.
This is opposed to only one, well defined, proton site, which should be reflected by one, relatively sharp, O-H stretch band. 
A peak fit analysis shows that we can reproduce the O-H stretch band by four Gaussian components at 3050, 3310, 3470, and 3540 \cmi (marked at I, II, IV, and V), and two Lorentzian components at 3410 and 3570 \cmi (marked as III and VI), hence suggesting that there are six distinctly different proton sites in the material.
 A comparison with Fig.~\ref{fig:Proton}(b) would suggest that bands II, III, IV and VI relate to the proton configuration according to \emph{type1}, \emph{i.e.} the \emph{Martinez1} structure, whereas bands I and V relate to the proton configuration according to \emph{type2}, \emph{i.e.} the \emph{Martinez2} structure.
  From the DFT calculations presented in Fig. \ref{fig:Proton} we also can see that O-H stretch vibrations involving H(1) protons are of higher frequency compared to the ones involving the H(2) proton. We thus attribute band III and IV to H(1) protons at the $32y$ position and band II to H(2) protons at the $4h$ position. 
 
\begin{figure}[t]
\begin{center}
\includegraphics[width=0.85\columnwidth]{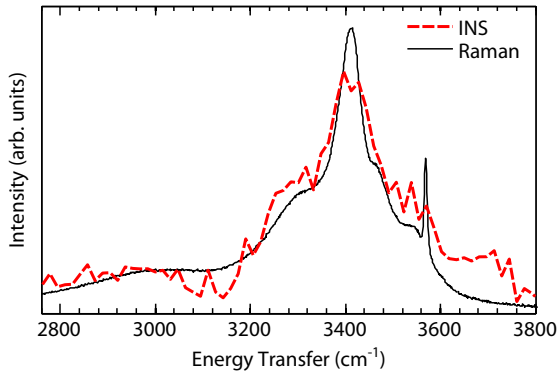}
\caption{Comparison between the O-H stretch region of the Raman and INS spectra of \BIOH.
The Raman spectrum was measured at room temperature, whereas the INS spectrum was measured at 10 K.
}
\label{fig:INS}
\end{center}
\end{figure}

Although the intensities of the different O-H stretch components provides a direct indication of the relative occupation of protons in the different sites, a quantitative assessment of the integrated intensity of the O-H stretch band/s is, generally, not straightforward, since the Raman scattering cross section may vary with the degree of hydrogen bonding, \textit{i.e.} the frequency of the vibration.
In order to elucidate the possible frequency dependency of the Raman scattering cross section, we also measured the O-H stretch region using INS, for which the intensity of a particular vibration is directly proportional to the number of vibrating species, irrespective of their vibrational frequency. 
A comparison of the Raman and INS spectra are shown in Fig.~\ref{fig:INS}. 
As can be seen, the shape of the O-H stretch band, measured with the two techniques, is indeed similar to each other, suggesting that there is no strong dependence of the Raman scattering cross section with the frequency of the O-H stretch vibrations in the material as studied here.

\begin{figure}[t]
\begin{center}
\includegraphics[width=0.98\columnwidth]{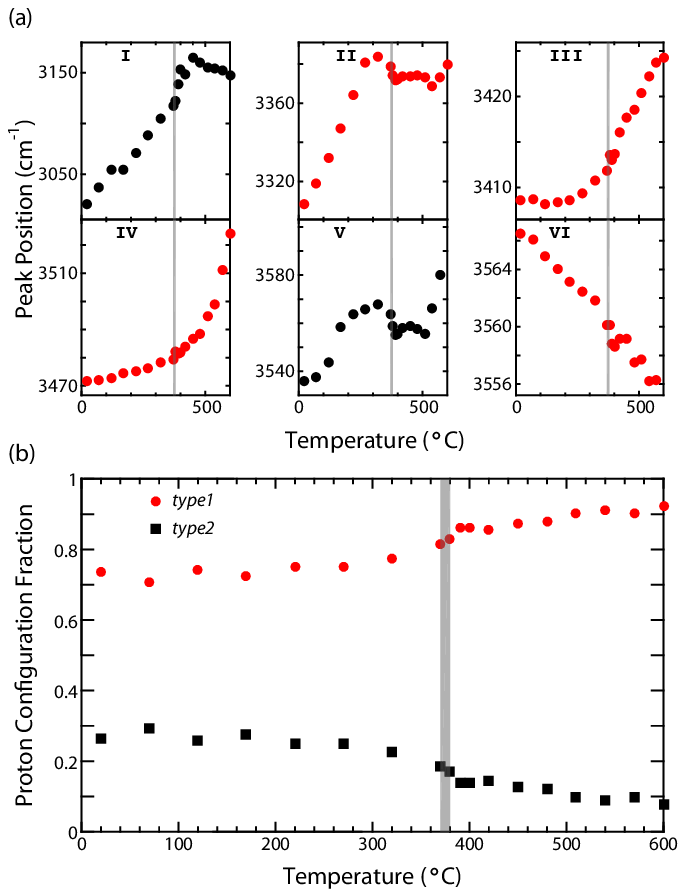}
\caption{
(Color online) 
(a) Temperature dependence of the peak position for the six O-H stretch bands, where bands II-IV and VI (in red color) relate to the proton configuration according to \emph{type1}, and where bands I and V (in black color) relate to the proton configuration according to \emph{type2}.
The shaded areas indicate the structural phase transition between the hydrated and intermediate phases. 
(b) Temperature dependence of the weights of the two phases, after normalization to unity.
}
\label{fig:OHBI}
\end{center}
\end{figure}

Further information about the proton sites can be obtained from the temperature dependence of the frequency of the different modes [Fig.~\ref{fig:OHBI}(a)].
Generally, an increase of temperature leads to a softening of vibrational modes due to thermal expansion of the chemical bonds involved in the vibrations. 
In the case of hydrogen bonding between the proton and a neighboring oxygen, however, the lattice expansion implies that the proton and neighboring oxygen are farther apart, which weakens the hydrogen bond and hence leads to an effective, counteracting, increase of the O-H stretch frequency.
With this in mind, we can draw the following conclusions based on our data in Fig.~\ref{fig:OHBI}(a).
The peak position of band II experiences a kink close to the hydrated-to-intermediate transition, implying a structural change in the tetrahedral layer of the $4h$ protons involved in the vibration. 
The $32y$ protons of band III are  not sensitive to this transition, by which we conclude that the structural change at the hydrated-to-intermediate transition is mainly restricted to the tetrahedral layers. 
Correspondingly, band I of the \emph{type2} phase is also due to $4h$ protons in the tetrahedral layer, while the pronounced anomaly of band V implies hydrogen bonds to the tetrahedral layer, perhaps from the apical O(2) oxygens as in the \emph{Martinez2} structure. 
Band IV behaves similar to band III with increasing temperature, and can thus also be assumed to belong to protons hydrogen bonded to the octahedral layer.
The continuous softening and narrow linewidth of band VI are both clear signs of a lack of hydrogen bonds. 
It is unclear how such a O-H group would fit into the structure and it is possible that band VI is due to trace amounts of impurities in the sample.
This is also consistent with its low intensity.

Analysis of the temperature dependence of the integrated areas of the different O-H stretch components shows how the fraction of the \emph{type1} and \emph{type2} phases changes with temperature, hence with the degree of hydration.
From Fig. \ref{fig:OHBI}(b), we observe that the fraction of \emph{type2} decreases with increasing temperature, suggesting that \emph{type1}, and in particular band III originating from the $32y$ protons [\emph{cf.} Fig. \ref{fig:Tstudy}(b)], is energetically more stable, although we cannot determine if this dependence is due mainly to transformation of \emph{type2} into \emph{type1} or if it is due to preferential dehydration for any of the two phases.
A plausible reason for the energetically higher stability of the \emph{type1} phase, may be that the non-centrosymmetric distortion of the In(1)O$_{6}$ octahedra creates a well-defined local energy minimum for the H(1) protons, which is in agreement with the relatively narrow linewidth of band II. 
This does not necessarily mean that the \emph{type1} phase reflects the global ground state, but perhaps a metastable state whose portion depends on how the sample is hydrated.

Finally, we note that the observation of two distinct proton configurations in \BIOH points towards differences in proton mobility between different proton sites.
In particular, we note that the energetically favored proton configurations, as featured by the full occupation of H(2) protons within the tetrahedral layer, and partial, but ordered, occupation of H(1) protons on the 32$y$ position, may indicate that the proton conductivity is higher in certain directions within the crystal structure. 
Moreover, the more well defined position on the $32y$ site in the \emph{type1} phase, together with it being more energetically stable at high temperatures, suggest that the  \emph{type2} phase might be favorable with regard to proton conductivity.
Although we are unable to determine precisely the factors determining the ratio of the two phases, our results provide some hints as to why the two phases can coexist.
On the one hand, we have a spectrally well-defined proton configuration, \emph{type1}, whereas on the other hand, we have the \emph{type2} configuration which is featured by generally broader spectroscopic features and thus a higher degree of structural variability in the material.
This may be indicative of a competition between energy and entropy at play, where the parameters of the hydration (\emph{e.g.} temperature, and time) may tip the balance of the two.
To this end, an investigation of the vibrational spectra as a function of systematic changes of the hydration conditions is likely to be beneficial for the clarification of the structure determining mechanisms involved, particularly if coupled to detailed studies of the dynamical behavior of protons, using \emph{e.g.} quasielastic neutron scattering.\cite{KAR15_QENSreview}

 \section{Conclusions}
To conclude, we find that the proton conducting oxide Ba$_{2}$In$_{2}$O$_{5}$(H$_{2}$O)$_{x}$, adopts three distinctly different (local) structures, depending on the level of hydration, $x$, and temperature, $T$. 
The structure evolves from a perovskite-like structure for the fully hydrated material ($x=1$) at $T=20$~$^{\circ}$C, with alternating layers of InO$_{6}$ octahedra, through a partially hydrated structure for 20~$^{\circ}$C~$<$~$T$~$<$~600~$^{\circ}$C, to a brownmillerite-type structured material, characterized by alternating layers of InO$_{6}$ octahedra and InO$_{4}$ tetrahedra, at essentially complete dehydration at 600~$^{\circ}$C. 
The transition from a perovskite-like to a brownmillerite-like structure is featured by a hydrated-to-intermediate phase transition at \emph{ca.} 370~$^{\circ}$C. 
The structure for the intermediate phase, is similar to the structure of the fully dehydrated material but with the difference that it is characterized by a non-centrosymmetric distortion of the InO$_{6}$ octahedra not present in the latter. 
Below the hydrated-to-intermediate phase transition, the dehydration is featured by the release of protons confined to the layers of InO$_{4}$ tetrahedra, whereas at  higher temperatures protons bound to oxygens of the layers of InO$_{6}$ are also released. 
Finally, we found that the O-H stretch region of the vibrational spectra is not consistent with a single-phase spectrum, but is in agreement with the intermixture of spectra associated with the lowest-energy and next-lowest-energy proton configurations in the structure of the material.
The amount of each phase is found to depend on how the material is hydrated.

\section*{Acknowledgements}
Funding from the Swedish Research Council (grant No. 2010-3519 and 2011-4887) is gratefully acknowledged. 
The STFC Rutherford Appleton Laboratory is thanked for access to neutron beam facilities.
We also thank S. M. H. Rahman at Chalmers University of Technology for the preparation of the sample.


\providecommand*{\mcitethebibliography}{\thebibliography}
\csname @ifundefined\endcsname{endmcitethebibliography}
{\let\endmcitethebibliography\endthebibliography}{}

\end{document}